\documentclass[onecolumn]{article}
\usepackage{amssymb}
\usepackage{amsmath}
\usepackage{geometry}
\usepackage{graphicx}
\usepackage{cite}
\usepackage{amsfonts}
\usepackage{color}

\begin{document}

\title{Graphene in curved Snyder space}
\author{B. Hamil$^{\ast }$ \\
D\'{e}partement de TC de SNV, Universit\'{e} Hassiba Benbouali, Chlef,
Algeria. \and H. Aounallah \\
Department of Science and Technology, Larbi Tebessi University, 12000
Tebessa, Algeria. \and B.C. L\" utf\"uo\u{g}lu \\
Department of Physics, Akdeniz University, Campus 07058, Antalya, Turkey, \\
Department of Physics, University of Hradec Kr\'alov\'e, Rokitansk\'eho 62,
500 03 Hradec Kr\'alov\'e, Czechia.}
\date{}
\maketitle

\begin{abstract}
The Snyder-de Sitter (SdS) model which is invariant under the action of the
de Sitter group, is an example of a noncommutative spacetime with three
fundamental scales. In this paper, we considered the massless Dirac fermions
in graphene layer in a curved Snyder spacetime which are subjected to an
external magnetic field. We employed representation in the momentum space to
derive the energy eigenvalues and the eigenfunctions of the system. Then, we
used the deduced energy function obtaining the internal energy, heat
capacity, and entropy functions. We investigated the role of the fundamental
scales on these thermal quantities of the graphene layer. We found that the
effect of the SdS model on the thermodynamic properties is significant.

\begin{description}
\item[Keyword:] Graphene; Snyder model; curved Snyder space; partition
function; thermodynamic functions.

\item[*Corresponding author, E-mail: ] hamilbilel@gmail.com
\end{description}
\end{abstract}

\section{Introduction}

Recently, an increasing interest is dedicated to the study of classical and
quantum mechanics on a quantized spacetime \cite{1,2,3,4,5,6,7,8,9,10}.
Historically Hartland S. Snyder was the pioneer of this idea. In 1947, he
proposed a fundamental length and stated noncommutative operators of the
quantized spacetime coordinates with four translation generators of the
algebra \cite{11}. Originally, this attempt was solely aimed to solve the
problems connected to UV divergences in quantum field theory. In the same
year he published a second article in which he discussed the effect of
quantized spacetime on the electromagnetic field theory \cite{12}. However,
this article was his last published article on this subject. Meanwhile, due
to the development of renormalization techniques, the UV divergences problem
in quantum field theory has resolved. For this reason, Snyder's idea was not
used more than some articles \cite{13,14,15,16,17,18}.
The curved
space selected by Snyder is the (3+1) de Sitter space, fabricated as the homogeneous space
\begin{equation}
dS^{(3+1)}=G/H=SO\left( 4,1\right) /SO\left( 3,1\right) ,
\end{equation}
where $SO(4,1)$ is the group of isometries, $H=SO(3,1)$ is
the Lorentz group while the Snyder-Galilean models are achieved as the
non-relativistic limit of the Snyder-Lorentzian models. The relativistic
modified quantum algebra proposed by Snyder is based on the following
commutation relations:
\begin{equation}
\left[ X_{\mu };P_{\nu }\right] =i\hbar \left( \eta _{\mu \nu }+\beta P_{\mu
}P_{\nu }\right) ;\text{ \ }\left[ X_{\mu };X_{\nu }\right] =i\hbar \beta
J_{\mu \nu };\text{ \ }\left[ P_{\mu };P_{\nu }\right] =0.
\end{equation}%
Here, $\eta _{\mu \nu }$ is the metric tensor where its signature is $\ %
\left[ \eta _{\mu \nu }\right] ={diag}\left(
\begin{array}{cccc}
-1 & 1 & 1 & 1%
\end{array}%
\right) $. $\beta $ is a coupling constant proportional to the Planck
length, $J_{\mu \nu }$ are the generators of Lorentz transformations while $%
\mu $ and $\nu $ are the spacetime indices with $\mu ,\nu =0,1,2,3$. Via the
presence of a fundamental constant, $\beta $, the model is seen to be an
example of doubly (or deformed) special relativity (DSR) \cite{19}.
Depending on the positive or negative values of the beta parameter, the
model is called as Snyder and anti-Snyder model, respectively \cite{2}.

In 1947, in order to make Snyder's theory invariant under the group of
translations that are based on the conformal group
$SO(1,5)$,  Yang extended Snyder's model to a de Sitter spacetime
background \cite{13}. The resulting model is characterized by three
invariant scales, the speed of light in vacuum, $c$, the Snyder parameter $%
\beta $, and the cosmological constant $\Lambda $. Therefore sometimes, that
model is preferred to be named as triply special relativity (TSR) instead of
the Snyder de Sitter (SdS) model. In momentum space a one-to-one
correspondence between the Snyder/anti-Snyder model and the de-Sitter/anti
de-Sitter spacetime is reported in \cite{20}.

The SdS model's algebra is constructed with the position $X_{\mu }$,
momentum $P_{\mu }$ and Lorentz generator $J_{\mu \nu }$ operators, which
obey the following algebra \cite{21,22}.
\begin{eqnarray}
\left[ J_{\mu \nu };X_{\sigma }\right] &=&i\hbar \left( \eta _{\mu \sigma
}X_{\nu }-\eta _{\nu \sigma }X_{\mu }\right) ;\text{ \ \ \ \ }\left[ J_{\mu
\nu };P_{\sigma }\right] =i\hbar \left( \eta _{\mu \sigma }P_{\nu }-\eta
_{\nu \sigma }P_{\mu }\right) ,  \notag \\
\left[ X_{\mu };P_{\nu }\right] &=&i\hbar \left( \eta_{\mu
\nu }+\alpha X_{\mu }X_{\nu }+\beta P_{\mu }P_{\nu }+\sqrt{\alpha \beta }%
\left( P_{\mu }X_{\nu }+X_{\nu }P_{\nu }-J_{\mu \nu }\right) \right) ,
\notag \\
\left[ X_{\mu };X_{\nu }\right] &=&i\beta \hbar J_{\mu \nu };\text{ \ \ \ \
\ \ \ \ \ \ \ \ }\left[ P_{\mu };P_{\nu }\right] =i\alpha \hbar J_{\mu \nu }.
\label{sds}
\end{eqnarray}%
This algebra can be regarded as a nonlinear realization of
Yang model. It should be noted that $\alpha $ and $\beta $ are the
coupling constants with dimension of inverse length and inverse mass,
they are defind as the square root of the cosmological constant $\sqrt{\alpha }\sim 10^{-24}cm^{-1}$ and with mass of Planck $\sqrt{\beta }\sim
10^{5}g^{-1}$ \cite{19}. In the limits $\alpha \rightarrow 0$, and $\beta
\rightarrow 0$ the algebra (\ref{sds}) reduces to the Snyder model in flat
space and to the de Sitter algebra, respectively \cite{23,24}.
The SdS phase space can
be realized in 6-dimensional space as $SO\left( 1,5\right) /SO\left( 1,3\right) \times O\left( 2\right) $  if $\alpha $, $\beta >0$ and $SO\left( 2,4\right) /SO\left( 1,3\right) \times O\left( 2\right) $ if $\alpha $, $\beta <0$  \cite{25}.
Recently, many authors have condensed their studies on the discussions over
the deformed canonical commutation relations \cite%
{25,26,27,28,29,30,31,32,33,34,35,36,37}.

In the last decade, Graphene has attracted the attention of theoretical and
experimental physicists\cite{g0,g1,g2}. A graphene material has a
two-dimensional structure that is constituted from carbon atoms in the
honeycomb lattice form which yields superior mechanical properties in
addition to optical and electronic properties \cite{g3,g4,g5,g6}. In the
theoretical perspective, a low energy excitation in a single graphene layer
is described by the massless Dirac equation \cite{g7,g8}.

In this paper, we consider massless fermions located in a graphene layer
under the effect of a perpendicular external magnetic field. We solve the
Dirac equation in the SdS model and obtain the energy eigenvalue function.
Then, we explore the thermodynamic functions in order to discuss the effects
of the fundamental scales of the SdS model. We present the manuscript as
follows: In section \ref{sec2} we introduce the SdS model briefly. In
section \ref{sec3}, we solve the massless Dirac equation. We obtain the
energy spectrum function analytically. Next, in section \ref{sec4} we define
the partition function. At the high-temperature limit, first, we derive the
internal energy functions, and then, the heat capacity and the entropy
functions. We demonstrate the thermodynamic functions versus temperature and
discuss the effect of the fundamental coupling constants of the SdS model.
We end the manuscript with a brief conclusion.

\section{Curved Snyder model}

\label{sec2}

In the nonrelativistic curved Snyder model, the modified commutation
relations between the position and momentum operators are given by \cite%
{22,37}:%
\begin{eqnarray}
\left[ X_{j};P_{k}\right] &=&i\hbar \left( \delta _{jk}+\alpha
X_{j}X_{k}+\beta P_{j}P_{k}+\sqrt{\alpha \beta }\left(
P_{j}X_{k}+X_{k}P_{j}\right) \right) ,  \notag \\
\left[ X_{j};X_{k}\right] &=&i\beta \hbar J_{jk},\text{ \ \ \ \ \ \ \ \ \ \
\ \ }\left[ P_{j};P_{k}\right] =i\alpha \hbar J_{jk},
\end{eqnarray}%
where $J_{jk}=\left( X_{j}P_{k}-X_{k}P_{j}\right)$. In
\cite{22,37}, one set of the $X_{j}$ and $P_{j}$ operators which satisfies
the algebra is expressed in the canonical coordinates as
\begin{eqnarray}
X_{j} &=&\mathcal{X}_{j}+\lambda \sqrt{\frac{\beta }{\alpha }}\mathcal{P}%
_{j}=i\hbar \sqrt{1-\beta p^{2}}\frac{\partial }{\partial p_{j}}+\lambda
\sqrt{\frac{\beta }{\alpha }}\frac{p_{j}}{\sqrt{1-\beta p^{2}}},
\label{CSM1} \\
P_{j} &=&-\sqrt{\frac{\alpha }{\beta }}\mathcal{X}_{j}+\left( 1-\lambda
\right) \mathcal{P}_{j}=-i\hbar \sqrt{\frac{\alpha }{\beta }}\sqrt{1-\beta
p^{2}}\frac{\partial }{\partial p_{j}}+\left( 1-\lambda \right) \frac{p_{j}}{%
\sqrt{1-\beta p^{2}}}.  \label{CSM2}
\end{eqnarray}%
Here, $\lambda $ is an arbitrary real parameter. Note that %
$p_{j}$ is bounded in the range of
$-\frac{1}{\sqrt{\beta }}<p_{j}<\frac{1}{\sqrt{\beta
}}$. If we consider a case, where $\left( \left\langle P_{j}\right\rangle
=\left\langle X_{j}\right\rangle =0\right) $, we obtain the uncertainty
relation as
\begin{equation}
\left( \Delta X\right) _{j}\left( \Delta P\right) _{k}\geq \frac{\hbar }{2}%
\left( \delta _{jk}+\alpha \left( \Delta X\right) _{j}\left( \Delta X\right)
_{k}+\beta \left( \Delta P\right) _{j}\left( \Delta P\right) _{k}-\sqrt{%
\alpha \beta }\left( \left( \Delta P\right) _{j}\left( \Delta X\right)
_{k}+\left( \Delta X\right) _{j}\left( \Delta P\right) _{k}\right) \right) .
\label{7}
\end{equation}%
In one dimension case, Eq. (\ref{7}) reduces to%
\begin{equation}
\left( \Delta X\right) \left( \Delta P\right) \geq \frac{\hbar }{2}\frac{%
\left( 1+\alpha \left( \Delta X\right) ^{2}+\beta \left( \Delta P\right)
^{2}\right) }{1+\hbar \sqrt{\alpha \beta }}.
\end{equation}%
As a conclusion, the modification in the algebra produces the following
minimal uncertainties in both position and momentum measurements
\begin{equation}
\left( \Delta X\right) _{\min }=\frac{\hbar \sqrt{\beta }}{\sqrt{1+2\hbar
\sqrt{\alpha \beta }}}\sim \hbar\sqrt{\beta};\text{ \ \ \
\ }\left( \Delta P\right) _{\min }=\frac{\hbar \sqrt{\alpha }}{\sqrt{%
1+2\hbar \sqrt{\alpha \beta }}}\sim \hbar\sqrt{\alpha}.
\end{equation}%
We note that, if $\alpha $, $\beta <0$ minimal uncertainties do not emerge
and all real values of $P_{i}$ are allowed. Before we proceed through the
next section, it is worth remarking the change of the definition of the
scalar product with the following form of \cite{22},
\begin{equation}
\left\langle \varphi \right. \left\vert \psi \right\rangle =\int \frac{d^{3}p%
}{\sqrt{1-\beta p^{2}}}\varphi ^{\ast }\left( p\right) \psi \left( p\right) .
\end{equation}

\section{Graphene in an external magnetic field}\label{sec3}

In this section, we solve the (2+1)-dimensional massless Dirac
equation in the curved Snyder model in the presence of the uniform magnetic
field $B$, which is directed along the $z$ axis. We assume $B>0$ and employ $%
A=\frac{B}{2}\left( -X_{2},X_{1}\right) $ gauge. We start with the Dirac
equation
\begin{equation}
i\hbar\frac{\partial }{\partial t}\Psi=H\Psi.
\end{equation}%
Here, $\psi $ is a two-dimensional wave function that describes the electron
states between the two Dirac points $A$ and $B$, while the
Dirac Hamiltonian is
\begin{equation}
H=V_{F}\overrightarrow{\mathbf{\alpha }}\cdot \left( \overrightarrow{P}-%
\frac{e}{c}\overrightarrow{A}\right) ,  \label{11}
\end{equation}%
where $V_{F}=\left( 1.12\pm 0.02\right) \times 10^{6} \frac{m}{s}^{-1}$
is the Fermi velocity.
We define a fundamental length scale,
$\ell_{B}$,  in the presence of an external magnetic field via
$\ell_{B}=\sqrt{\frac{\hbar c}{eB}}$, and express the Hamiltonian in the
matrix form
\begin{equation}
H=\left(
\begin{array}{cc}
H_{A} & 0 \\
0 & H_{B}%
\end{array}%
\right) ,
\end{equation}%
with two Hamiltonian operators for the two Dirac points $A$ and $B$ as
\begin{eqnarray}
H_{A} &=&V_{F}\left(
\begin{array}{cc}
0 & \left( P_{1}-iP_{2}\right) +\frac{\hbar}{2\ell_B^2}%
\left( X_{2}+iX_{2}\right) \\
\left( P_{1}+iP_{2}\right) +\frac{\hbar}{2\ell_B^2}\left(
X_{2}-iX_{2}\right) & 0%
\end{array}%
\right) , \\
H_{B} &=&V_{F}\left(
\begin{array}{cc}
0 & \left( P_{1}+iP_{2}\right) +\frac{\hbar}{2\ell_B^2}%
\left( X_{2}-iX_{2}\right) \\
\left( P_{1}-iP_{2}\right) +\frac{\hbar}{2\ell_B^2}\left(
X_{2}+iX_{2}\right) & 0%
\end{array}%
\right).
\end{eqnarray}%
We write the two-component wave function as
\begin{equation}
\Psi=e^{\frac{iE}{\hbar}t}\left(
\begin{array}{c}
\psi ^{A} \\
\psi ^{B}%
\end{array}%
\right) ,
\end{equation}%
where $\psi ^{A}$ and $\psi ^{B}$ are two
dimensional eigenstates. To obtain the energy eigenvalue
equation of the wave function at the Dirac point $A$, we solve the following
system of two coupled equations:
\begin{eqnarray}
\left[ P_{1}-iP_{2}+\frac{\hbar}{2\ell_B^2}\left(
X_{2}+iX_{1}\right) \right] \psi ^{B} &=&\frac{E}{V_{F}}%
\psi ^{A}, \\
\left[ P_{1}+iP_{2}+\frac{\hbar}{2\ell_B^2}\left(
X_{2}-iX_{1}\right) \right] \psi ^{A} &=&\frac{E}{V_{F}}%
\psi ^{B }.
\end{eqnarray}%
Out of the coupled system, we obtain the following decoupled differential
equation for the component $\psi ^{A}$.
\begin{eqnarray}
\Bigg[ P_{1}^{2}+P_{2}^{2}+\left(\frac{\hbar}{2\ell_B^{2}}\right)^2\left(
X_{2}^{2}+X_{1}^{2}\right) +i\left[ P_{1};P_{2}\right] +i\left(\frac{\hbar}{2\ell_B^{2}}\right)^2\left[ X_{1};X_{2}\right]+\frac{\hbar}{2\ell_B^2}\left( P_{1}X_{2}+X_{2}P_{1}-X_{1}P_{2}-P_{2}X_{1}\right)  \nonumber \\
+ \frac{i \hbar}{2\ell_B^2}\left[ X_{1};P_{1}\right] +\frac{i \hbar}{2\ell_B^2}\left[ X_{2};P_{2}\right]\Bigg] \psi ^{A}=\frac{E^{2}}{V_{F}^{2}}\psi ^{A}.
\end{eqnarray} %
Then, we employ the position and momentum operators given in Eqs. (\ref{CSM1}%
) and (\ref{CSM2}). We find
\begin{eqnarray}
\Bigg\{
\left[\frac{\beta }{\alpha } \left(\frac{\hbar}{2\ell_B^{2}}\right)^2+1\right]
\left(\sqrt{\frac{\alpha }{\beta }}\mathcal{X}_{j}\right)^{2}+\left( 1-2\lambda +\lambda^2 \left[\frac{\beta }{\alpha } \left(\frac{\hbar}{2\ell_B^{2}}\right)^2+1\right] \right) \mathcal{P}_{j}^{2}-\left[\frac{\beta }{\alpha } \left(\frac{\hbar}{2\ell_B^{2}}\right)^2+1\right] \alpha \hbar L_{z} \nonumber  \\
-\left(1- \lambda \left[\frac{\beta }{\alpha } \left(\frac{\hbar}{2\ell_B^{2}}\right)^2+1\right] \right)\sqrt{\frac{\alpha}{\beta}}\left( \mathcal{X}_{j}\mathcal{P}_{j}+\mathcal{P}_{j}\mathcal{X}_{j}\right)
-\frac{\hbar^2}{\ell_B^{2}}\left( 1 +\frac{L_{z}}{\hbar}+\frac{\beta}{2} \mathcal{P}_{j}^{2}\right)
\Bigg\} \psi ^{A}=\frac{E^{2}}{V_{F}^{2}}\psi ^{A}.  \label{19}
\end{eqnarray}%
%
In order to simplify the differential equation we assume
\begin{eqnarray}
  \lambda &=& \bigg[\frac{\beta}{\alpha}\left(\frac{\hbar}{2\ell_B^2}\right)^2+1\bigg]^{-1}
\end{eqnarray}
Then, $\left( \mathcal{X}_{j}\mathcal{P}_{j}+\mathcal{P}_{j}\mathcal{X}%
_{j}\right) $ term vanishes and Eq. (\ref{19}) becomes%
\begin{equation}
\left\{ -\left( 1-\beta p^{2}\right) \left( \frac{\partial ^{2}}{\partial
p^{2}}+\frac{1}{p}\frac{\partial }{\partial p}-\frac{\mu ^{2}}{p^{2}}\right)
+\beta p\frac{\partial }{\partial p}+\frac{1}{2\ell_\mathcal{B}^{2}}\left(
\frac{1}{2\ell_\mathcal{B}^{2}}-1\right) \frac{\beta ^{2}p^{2}}{1-\beta p^{2}}-\beta \mu -\frac{\beta }{\ell_\mathcal{B}^{2}}\left( \mu +1\right) \right\}
\Phi ^{A}=\frac{\beta \mathcal{E}^{2}}{\hbar ^{2}V_{F}^{2}}\Phi ^{A},
\label{21}
\end{equation}%
%
where%
\begin{equation}
\psi ^{A }=\frac{e^{i\mu \varphi }}{\sqrt{2\pi }}%
\Phi ^{A };\text{ \ \ }\left( \mu =0,\pm 1,\pm 2,...\right)
,\text{\ }
\end{equation}%
and%
\begin{equation}
\frac{1}{\ell_\mathcal{B}^{2}=\frac{\alpha}{\lambda}%
\frac{1}{\ell_{B}^{2}}};\text{ \ \ }\mathcal{B}\equiv \frac{\lambda B}{%
\alpha };\text{ \ \ }\mathcal{E}^{2}\equiv \frac{\lambda E^{2}}{\alpha }.
\end{equation}%
We perform a change of the variable%
\begin{equation}
\rho =\frac{\sin ^{-1}\sqrt{\beta }p}{\sqrt{\beta }},
\end{equation}%
which maps the allowed range from $\frac{-1}{\sqrt{\beta }}<p<\frac{1}{\sqrt{%
\beta }}$ to $\frac{-\pi }{2\sqrt{\beta }}<\rho <\frac{\pi }{2\sqrt{\beta }}$%
. Eq. (\ref{21}) reduces to the form of
\begin{equation}
\left[ \frac{\partial ^{2}}{\partial \rho ^{2}}+\sqrt{\beta }\cot \left(
\sqrt{\beta }\rho \right) \frac{\partial }{\partial \rho }-\beta \mu
^{2}\cot ^{2}\left( \sqrt{\beta }\rho \right) - \frac{\beta}{2\ell_\mathcal{B}^{2}}
\left( \frac{1}{2 \ell_\mathcal{B}^{2}}-1\right) \tan ^{2}\left( \sqrt{\beta }\rho \right) +\beta \mu +\frac{\beta \left( \mu +1\right)
}{\ell_\mathcal{B}^{2}}+\frac{\beta \mathcal{E}^{2}}{\hbar ^{2}V_{F}^{2}}\right] \Phi
^{A}=0.  \label{can}
\end{equation}%
%
%
By ansatz, we assume $\Phi ^{A }=\sin ^{\mu }\left( \sqrt{%
\beta }\rho \right) \cos ^{\delta }\left( \sqrt{\beta }\rho \right) \digamma
$. Then, Eq. (\ref{can}) becomes
\begin{eqnarray}
\Bigg\{
\frac{\partial ^{2}}{\partial \rho ^{2}}+\sqrt{\beta }\left[ \left( 1+2\mu
\right) \cot \left( \sqrt{\beta }\rho \right) -2\delta \tan \left( \sqrt{\beta }\rho \right) \right] \frac{\partial }{\partial \rho }+\beta \left[ \delta \left( \delta -1\right) -\frac{1}{2\ell_\mathcal{B}^{2}}\left( \frac{1}{2\ell_\mathcal{B}^{2}}-1\right) \right] \tan ^{2}\left( \sqrt{\beta }\rho \right) \nonumber \\
+ \left(\frac{1}{\ell_\mathcal{B}^{2}}-2\delta  \right)\beta \left(\mu +1\right)  +\frac{\beta \mathcal{E}^{2}}{\hbar ^{2}V_{F}^{2}}\Bigg\} \digamma =0.  \label{26}
\end{eqnarray}%
%
We fix the parameter $\delta $ by requiring the coefficient of the $\tan
^{2}\left( \sqrt{\beta }\rho \right) $ to vanish:%
\begin{equation}
\delta \left( \delta -1\right) -\frac{1}{2\ell _{\mathcal{B}}^{2}}\left(
\frac{1}{2\ell _{\mathcal{B}}^{2}}-1\right) =0.
\end{equation}%
We determine the roots of the quadratic equation of $\delta $ as%
\begin{equation}
\delta =\frac{1}{2\ell _{\mathcal{B}}^{2}};\text{ }\delta ^{\prime }=1-\frac{%
1}{2\ell _{\mathcal{B}}^{2}}.  \label{29}
\end{equation}%
We note that the second root does not lead to a physically acceptable wave function unless we impose the condition ${\ell_\mathcal{B}^{2}} > 1/2$.
However, this condition is valid only for small enough field strengths.
Therefore, we use the first root in Eq.~(\ref{29}). We find the reduced
equation in the form of
\begin{equation}
\left[ \frac{\partial ^{2}}{\partial \rho ^{2}}+\sqrt{\beta }\left[ \left(
1+2\mu \right) \cot \left( \sqrt{\beta }\rho \right) -2\delta \tan \left(
\sqrt{\beta }\rho \right) \right] \frac{\partial }{\partial \rho }+\frac{%
\beta \mathcal{E}^{2}}{\hbar ^{2}V_{F}^{2}}\right] \digamma =0.  \label{30}
\end{equation}%
We define a new variable $q=\sin ^{2}\left( \sqrt{\beta }\rho \right) $.
Then, Eq. (\ref{30}) turns into the form of the hypergeometric equation \cite%
{38}.
\begin{equation}
\left[ q\left( 1-q\right) \frac{\partial ^{2}}{\partial q^{2}}+\left[ \left(
1+\mu \right) -\left( \frac{3}{2}+\mu +\delta \right) q\right] \frac{%
\partial }{\partial q}+\frac{\mathcal{E}^{2}}{4\hbar ^{2}V_{F}^{2}}\right]
\digamma =0.
\end{equation}%
This equation has a regular solution at $q=0$ that is written in terms of
hypergeometric functions as
\begin{equation}
\digamma =\mathbf{F}\left( a,b,1+\mu ;q\right) ,
\end{equation}%
with the following parameters:
\begin{eqnarray}
a &=&\frac{1}{4}+\frac{\delta +\mu }{2}+\frac{1}{2}\sqrt{\frac{1}{4}+\mu
\left( 1+\mu \right) +2\mu \delta +\delta \left( \delta +1\right) +\frac{%
\mathcal{E}^{2}}{\hbar ^{2}V_{F}^{2}}}, \\
b &=&\frac{1}{4}+\frac{\delta +\mu }{2}-\frac{1}{2}\sqrt{\frac{1}{4}+\mu
\left( 1+\mu \right) +2\mu \delta +\delta \left( \delta +1\right) +\frac{%
\mathcal{E}^{2}}{\hbar ^{2}V_{F}^{2}}}.
\end{eqnarray}
%
The hypergeometric function $\mathbf{F}\left( a,b,c
;q\right) $ is determined by the hypergeometric series \cite{38} as follows:
\begin{equation}
\mathbf{F}\left( a,b,c,q\right) =\sum_{n=0}\frac{\left( a\right) _{n}\left(
b\right) _{n}}{\left( c\right) _{n}}\frac{q^{n}}{n!},
\end{equation}
where the parameters of the hypergeometric series are given
by
\begin{equation}
\left( a\right) _{n}=\frac{\Gamma \left( a+n\right) }{\Gamma \left( a\right)};
\qquad \left( b\right) _{n}=\frac{\Gamma \left( b+n\right) }{\Gamma \left(b\right) };
\qquad\left( c\right) _{n}=\frac{\Gamma \left( c+n\right) }{\Gamma
\left( c\right) }.
\end{equation}
The series reduces to a polynomial if $a$ or $b$ is a
negative integer. Using the expressions of $b,$ we finally obtain
\begin{equation}
E_{n}=\pm \hbar V_{F}\sqrt{4\theta n^{2}+2\theta n\left( 1+\frac{1}{\ell
_{B}^{2}\theta }+2\mu \right) },  \label{37}
\end{equation}
where $\theta =\alpha +\frac{\hbar ^{2}\beta }{4\ell _{B}^{4}}$.
We would like to emphasize that the latter equation represents the main
result of our paper. We observe that the introduced deformed Heisenberg
algebra has influence on results. We also note that the energy spectrum
depends on $n^{2}$, which is a feature of hard confinement.
Furthermore, the energy spectrum values increase
proportional to $n$ for the large quantum number $n$.
It should be pointed out that for $n=0$, the energy level $E_{n}=0$, which means  the energy level at higher levels ($n=1,2,...$)
are distributed symmetrically around $n=0$. In addition, the energy level is proportional to $\sqrt{n^{2}}$, which implies the energy spacing between adjacent levels is not constant. For large $n$ the energy spacing becomes constant
\begin{equation}
\lim_{n\rightarrow \infty }\Delta E_{n}=\left\vert E_{n+1}-E_{n}\right\vert
=\hbar \omega _{c},
\end{equation}
where  $\omega _{c}=2V_{F}\sqrt{\theta }$  can be interpreted
as classical cyclotron frequency. As $\alpha$ and $\beta$ are small in
comparison with the other quantities in the theory, we expand (\ref{37}) to
first order in $\alpha$ and $\beta$, we obtain
\begin{equation}
E_{n}=\pm \frac{\hbar V_{F}}{\ell _{B}}\sqrt{2n}\pm \theta \hbar V_{F}\sqrt{2n\left( n+\mu +\frac{1}{2}\right) ^{2}}. \label{Ee}
\end{equation}
Here, the first term represent the Landau levels of electrons in
graphene while the second term is the quantum gravity correction. Now, let
us consider the following particular cases.
\begin{enumerate}
\item In the limit $\beta \rightarrow 0$, we recover the results for anti-de
Sitter space \cite{48}.%
\begin{equation}
E_{n}=\pm \hbar V_{F}\sqrt{4\alpha n^{2}+2\alpha n\left( 1+2\mu \right) +%
\frac{2n}{\ell _{B}^{2}}}.
\end{equation}

\item In the limit $\alpha \rightarrow 0$, we obtain the energy spectrum in
the Snyder space.
\begin{equation}
E_{n}=\pm \hbar V_{F}\sqrt{\frac{\hbar ^{2}\beta }{\ell _{B}^{4}}n^{2}+\frac{%
\hbar ^{2}\beta }{2\ell _{B}^{4}}n\left( 1+2\mu \right) +\frac{2n}{%
\ell_{B}^{2}}}.
\end{equation}

\item
In anti-Snyder de Sitter model where $\alpha <0$ and
$\beta <0$, the energy spectrum is,
\begin{equation}
E_{n}=\pm \hbar V_{F}\sqrt{\frac{2n}{\ell _{B}^{2}}-\left( \alpha +\frac{%
\hbar ^{2}\beta }{4\ell _{B}^{4}}\right) \left[ 4n^{2}+2n\left( 1+2\mu
\right) \right] }.
\end{equation}%
in this case the energy spectrum $E_{n}$ becomes complex
when the quantum number $n$ is large. This stipulates an upper bound on the allowed values of $n$.

\item In the limit $\beta \rightarrow 0$ and $\alpha \rightarrow 0$, we get
the ordinary quantum mechanical result \cite{49,50}.
\begin{equation}
E_{n}=\pm \frac{\hbar V_{F}}{\ell _{B}}\sqrt{2n}.
\end{equation}

\item
If $\delta =1-\frac{1}{2\ell _{\mathcal{B}}^{2}}$, the
energy spectrum is,
\begin{equation}
E_{n}=\pm \hbar V_{F}\sqrt{\theta }\sqrt{4n^{2}+4n\mu +6n+2\mu +2-\left(
1+2\mu +2n\right) \frac{e\mathcal{B}}{\hbar c}}.
\end{equation}
\end{enumerate}
On an other side, if $a$ or $b$ is a nonnegative integer,
the hypergeometric series converges absolutely for all values of $\left\vert
q\right\vert <1$ \cite{51} and,
\begin{equation}
\frac{\left( a\right) _{n}\left( b\right) _{n}}{\left( c\right) _{n}}=\frac{\Gamma \left( c\right) }{\Gamma \left( a\right) \Gamma \left( b\right) }n^{\delta -\frac{3}{2}}\left[ 1+\mathcal{O}\left( n^{-1}\right) \right] .
\end{equation}%
For to make the hypergeometric function to be regular
at $q=1$, we must impose $\ell _{\mathcal{B}}^{2}>1$ , which is valid
only for small enough field strengths. In the case where the particles cannot be
bounded, only scattering solutions occur and the energy spectrum becomes
continuous.

\begin{figure}[tbh]
\centering
\includegraphics[width=0.5\linewidth]{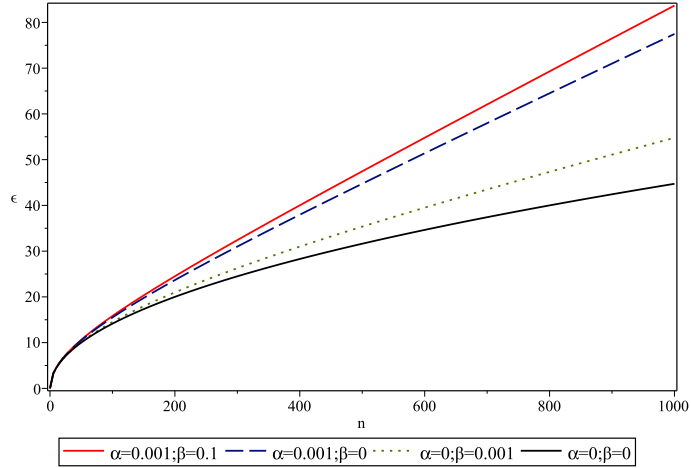}
\caption{$\frac{E_{n,\protect\mu =0}}{\hbar V_{F}}$ versus the quantum
number $n$ for different values of the deformation parameters.}
\label{fig1}
\end{figure}

In order to demonstrate the influence of the modified algebra on the energy
levels, we plot the energy levels $\epsilon =\frac{E_{n,\mu =0}}{\hbar V_{F}}
$ versus the quantum number, $n$, by employing different values of the
deformation parameters in Fig. \ref{fig1}. We observe that the contribution
of the $\alpha $ parameter is more significant than the $\beta$ parameter.

It should be noted that the dependence of the energy levels on SdS
parameters is only through $\theta $, this can be easily quantified. For
weak magnetic fields, the parameter $\theta $ identified as the cosmological
constant, $\theta \sim \alpha \sim 10^{-48}$. If the magnetic field is
extremely strong the parameter $\theta $ take the value $\theta \sim \frac{e^{2}B^{2}\beta }{c^{2}}\sim 10^{-44}B^{2}$.

\section{Thermodynamic functions}\label{sec4}

It is a well-known fact that an electron gas obeys the Fermi-Dirac quantum
statistic. However, in high temperatures or with the consideration of the
electron gas in a low density, the Maxwell-Boltzmann statistic can be used
instead \cite{52}. In this section we aim to determine the thermodynamic
properties of the of the graphene under a magnetic field in SdS space. We
suppose only fermions with positive energy ($E\geqslant 0$) constitute the
thermodynamic ensemble. Since we ignore the particle-particle interactions,
we take neither negative-energy excited states nor the phenomenon of
creation of particles into account \cite{50,53}. Therefore, we assume the
partition function contains only a sum over positive-energy states. We note
that this is an enormous simplification characteristic. {We begin by
computing the partition function of a single particle of the system, $Z$,
for the fixed angular momentum $\left( \mu =0\right)$:
\begin{equation}
Z=\sum_{n=0}^{+\infty }e^{-\frac{E_{n}}{KT}}.  \label{46}
\end{equation}}%
Here, $K$ denotes the Boltzmann constant, $T$ represents the thermodynamic
temperature and $E_{n}$ is the energy eigenvalues. {red}{We
use the derived energy eigenvalue function given in Eq. \eqref{Ee} in Eq. \eqref{46}.
We obtain the partition function in the form of
\begin{equation}
Z=\sum_{n=0}^{+\infty }e^{-\frac{\hbar V_{F}}{KT}\sqrt{2\theta \left(
2n^{2}+1\right) +\frac{2n}{\ell _{B}^{2}}}}.  \label{47}
\end{equation}
Since $\alpha $ and $\beta $ are small in comparison with the other quantities
in the theory, we expand Eq. (\ref{47}) till to the first order of $\alpha $ and $\beta $.
We obtain%
\begin{equation}
Z=\sum_{n=0}^{+\infty }\left[ 1-\frac{\hbar V_{F}\theta }{2}\sqrt{2n}-\frac{%
\hbar V_{F}\theta }{4}\left( 2n\right) ^{\frac{3}{2}}\right] e^{-\overline{%
\beta }\sqrt{n}},  \label{48}
\end{equation}
where $\overline{\beta }=\frac{1}{\tau },$ and $\tau$ is the reduced temperature defined with
\begin{equation}
\tau =\frac{KT\ell _{B}}{\hbar V_{F}\sqrt{2}}=\frac{T}{T_{0}}.
\end{equation}
Here, $T_{0}=\frac{\hbar V_{F}\sqrt{2}}{K\ell _{B}}$
is the temperature reference value, for instance when $B=18T$, the value of this
temperature becomes $T_{0}=3551 {K}$. It is worth noting that the first term in Eq. \eqref{48} is the ordinary
partition function of the graphene under a magnetic field \cite{54}%
\begin{equation}
Z_{0}=\sum_{n=0}^{+\infty }e^{-\overline{\beta }\sqrt{n}}=\frac{1}{\overline{%
\beta }^{2}}+\zeta \left( 0\right) =\frac{1}{\overline{\beta }^{2}}-\frac{1}{%
2}.  \label{50}
\end{equation}
We calculate the second and third terms of Eq. (\ref{48}) by using the
derivatives of Eq. (\ref{50}) as follows:%
\begin{equation}
\sum_{n=0}^{+\infty }\left[ \frac{\hbar V_{F}\theta }{2}\sqrt{2n}+\frac{%
\hbar V_{F}\theta }{4}\left( 2n\right) ^{\frac{3}{2}}\right] e^{-\overline{%
\beta }\sqrt{n}}=\left[ -\frac{\hbar V_{F}\theta \sqrt{2}}{2}\frac{\partial
}{\partial \overline{\beta }}-\hbar V_{F}\theta \sqrt{2}\frac{\partial ^{3}}{%
\partial \overline{\beta }^{3}}\right] Z_{0}.
\end{equation}
After all, we obtain the total partition function of the system in the SdS space in the form of%
\begin{equation}
Z=\tau ^{2}-\frac{1}{2}-\hbar V_{F}\theta \sqrt{2}\tau ^{3}\left( 1+24\tau
^{2}\right).
\end{equation}
Next, we derive the thermal properties of our system, such as the internal
energy and the specific heat through the numerical partition function $Z$
via the following relations:
\begin{equation}
U=\frac{\ell _{B}}{\hbar V_{F}\sqrt{2}}\mathbf{U}=\tau ^{2}\frac{\partial }{%
\partial \tau }\ln Z,\qquad C=\frac{\partial U}{\partial \tau }.
\end{equation}%
We take $\ell _{B}=\hbar =c=K=1$, and demonstrate all profiles of the thermodynamic quantities
as a function of $\tau $ for various values of the SdS parameters.}

\begin{figure}[tbh]
\centering
\includegraphics[width=0.5\linewidth]{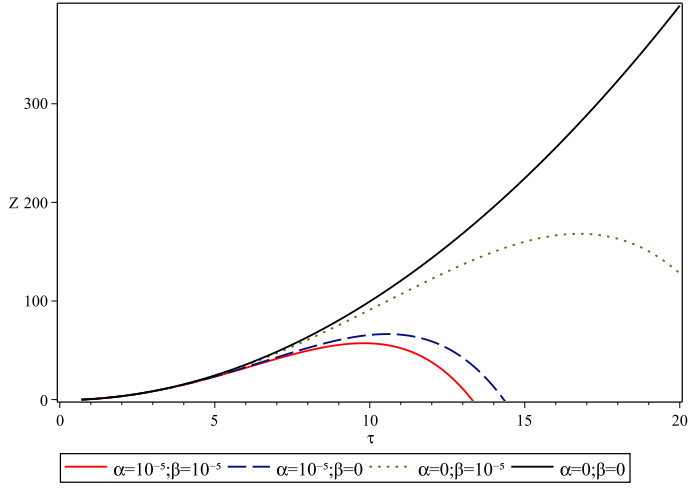}
\caption{Partition function versus the reduced temperature.}
\label{fig2}
\end{figure}

First, we plot the partition function versus $\tau $ in Fig. (\ref{fig2}). We observe
a monotonic increase in the partition function in the ordinary quantum
mechanic limit. This characteristic behavior drastically changes in the
existence of the SdS model parameters. We observe a decrease in the partition function
while the temperature increases at the high-temperature values. The amount
of the decrease in the partition function value increases when de Sitter
spacetime is taken into account instead of the Snyder model.

{We present the characteristic behavior of the internal energy function versus the dimensionless reduced temperature for different values of the SdS parameters in Fig. \eqref{fig3}. In the ordinary quantum mechanic limit, we observe a linear increase. When we consider a comparison between the role of the and parameters, we observe that the internal energy is being modified significantly in the de Sitter space, rather than the Snyder model because of the dependence on the strength of the magnetic field. We also see that in the vicinity of zero, there is no difference between the standard and the modified internal energy, which implies that the effects of quantum gravity become more obvious only at high temperatures.
\begin{figure}[tbh]
\centering
\includegraphics[width=0.5\linewidth]{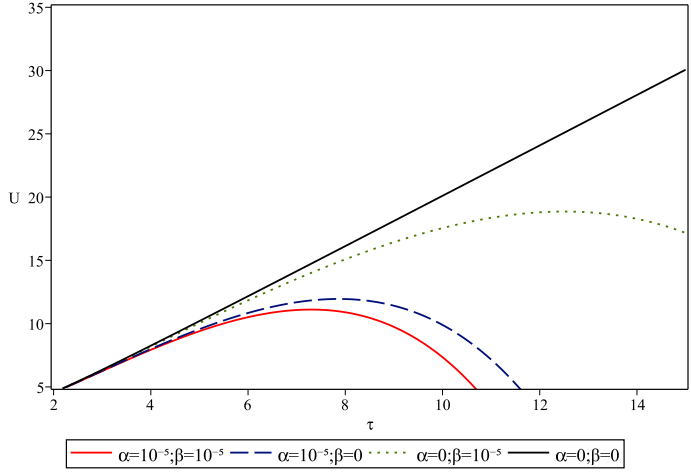}
\caption{Internal energy versus the reduced temperature.}
\label{fig3}
\end{figure}}

{Finally, we illustrate the heat capacity function versus the reduced temperature in
Fig. \eqref{fig4} by considering different values of the SdS parameters. In
the ordinary quantum mechanic limit, we observe that the heat capacity will
tend to a constant value at high temperature. We also see a decrease in the
heat capacity function for high-temperature in the existence of $\left(
\alpha ,\beta \right) $. When we consider a comparison in between the
parameters, like the other cases, we realize that the role of the parameter
$\alpha  $ is more significant than the $\beta$ parameter.
\begin{figure}[tbh]
\centering
\includegraphics[width=0.5\linewidth]{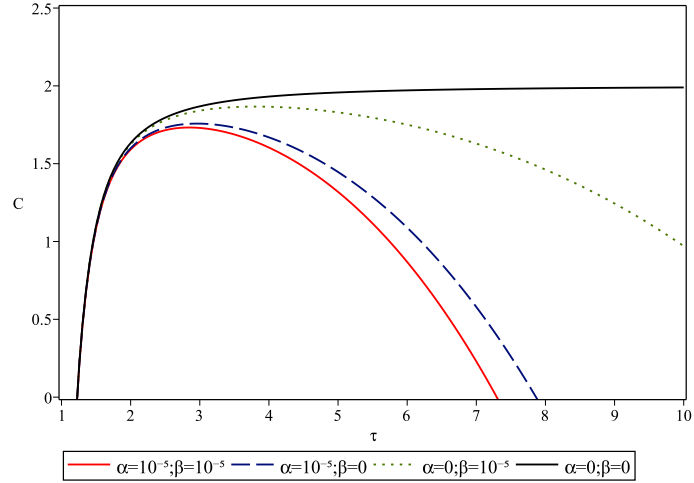}
\caption{Specific heat function versus the reduced temperature.}
\label{fig4}
\end{figure}}

It is worthwhile to note that all thermodynamic quantities obtained
numerically in our work show that the effects of the curved Snyder model on
the statistical properties of graphene are important only in the
high-temperature regime, contrary the case at low temperatures. The effect
of the SdS model becomes insignificant and the curves join rapidly as the
temperature decreases. Our conclusion that the quantum gravity effects have
concrete effects specifically at high temperature limits.

Finally, when SdS parameters $\alpha =\beta =0$ our results agrees exactly
with that of \cite{49,54}. One of the biggest issues in physics at present
is to combine the quantum theory and the theory of general relativity into a
unified framework, different approaches towards such a theory of quantum
gravity have been elaborated. Despite that, one major obstacle is the
absence of experimental confirmation of quantum gravitational effects \cite{55}.
The results we presented here may afford a source of information to
probing Planck-scale physics in future experiments.

\section{Conclusion}\label{conc}

In this paper, we considered a graphene layer which is under the influence
of an external magnetic field. We assumed the applied field to be
perpendicular to the layer and solved the massless Dirac equation in the
(2+1) dimension in the SdS model. We derived an analytic solution to the
wave and energy eigenvalue functions. The essential characteristic of energy
levels is the existence of zero-energy states.

Then, we investigated some of the statistical characteristics of the
considered system at high-temperatures by comparison of the thermodynamic
functions. We found that the fundamental scales of the model have an
important role on the thermal quantities. We comprehended that the
contribution of the deformation parameter of the de Sitter spacetime is more
significant than the Snyder model parameter.  We also found the influence of
SdS model can be seen only at high temperatures limits.

\section*{Acknowledgment}
The authors thank the referees for a thorough reading of our manuscript and
for constructive suggestion.


%
%


\begin{thebibliography}{99}
\bibitem{1} S. Mignemi, Phys. Rev. D \textbf{84}, 025021 (2011).

\bibitem{2} S. Mignemi, Journal of Physics: Conference Series \textbf{343},
012074 (2012).

\bibitem{3} S. A. Franchino-Vinas and S. Mignemi, Phys. Rev. D
\textbf{98}, 065010 (2018).

\bibitem{4} S. Meljanac, S. Mignemi, J. Trampetic and J. You, Phys. Rev. D
\textbf{96}, 045021 (2017).

\bibitem{5} S. Mignemi, Ukr. J. Phys. \textbf{64}, 991 (2019).

\bibitem{6} S. A. Franchino-Vinas, S. Mignemi, arXiv:2005.12610.

\bibitem{7} L. Lu, A. Stern, Nucl. Phys. B \textbf{854}, 894 (2012).

\bibitem{8} W. S. Chung and H. Hassanabadi, Int. J. Theor. Phys. \textbf{58}%
, 2267 (2019).

\bibitem{9} M. V. Battisti and S. Meljanac, Phys. Rev. D \textbf{79}, 067505
(2009).

\bibitem{10} M. V. Battisti and S. Meljanac, Phys. Rev. D \textbf{82},
024028 (2010).

\bibitem{11} H. S. Snyder, Phys. Rev. \textbf{71}, 38 (1947).

\bibitem{12} H. S. Snyder, Phys. Rev. \textbf{72}, 68 (1947).

\bibitem{13} C. N. Yang, Phys. Rev. \textbf{72}, 874 (1947).

\bibitem{14} Y. A. Gol'fand, Sov. Phys. JETP \textbf{17}, 842 (1963).

\bibitem{15} V. G. Kadyshevsky, Sov. Phys. JETP \textbf{14}, 1340 (1962).

\bibitem{16} R. M. Mir-Kasimov, Sov. Phys. JETP \textbf{22}, 629 (1966).

\bibitem{17} Y. A. Gol'fand, Sov. Phys. JETP \textbf{16}, 184 (1963).

\bibitem{18} R. M. Mir-Kasimov, Sov. Phys. JETP \textbf{25}, 348 (1967).

\bibitem{19} G. Amelino-Camelia, L. Smolin, and A. Starodubtsev, Class.
Quant. Grav. \textbf{21}, 3095 (2004).

\bibitem{20} H. -Y. Guo, Yu. C.-G.Huang, Z. X. Tian, B. Zhou, Front. Phys.
China \textbf{2}, 358 (2007).

\bibitem{21} J. Kowalski-Glikman and L. Smolin, Phys. Rev. D \textbf{70},
065020 (2004).

\bibitem{22} S. Mignemi, Class. Quantum Grav. \textbf{29}, 215019 (2012).

\bibitem{23} S. Mignemi, Mod. Phys. Lett. A \textbf{20}, 1697 (2010).

\bibitem{24} W. S. Chung and H. Hassanabadi, Mod. Phys. Lett. A \textbf{26},
1750138 (2017).

\bibitem{25} S. Mignemi, Ukr. J. Phys. \textbf{64}, 991 (2019).

\bibitem{26} B. Ivetic, S. Meljanac, S. Mignemi, Class. Quantum Grav.
\textbf{31}, 105010 (2014).

\bibitem{27} S. A. Franchino-Vi\~{n}as, S. Mignemi, arXiv:1912.10962.

\bibitem{28} S. Mignemi, Annal. Phys. \textbf{522}, 924 (2010).

\bibitem{29} S. Mignemi, Class. Quant. Gravit. \textbf{26}, 245020 (2009).

\bibitem{30} S. Mignemi, R. Strajn, Adv. High Energy Phys. \textbf{2016},
1328284 (2016).

\bibitem{31} B. Hamil, M. Merad and T. Birkandan, Int. J. Mod. Phys. A
\textbf{35}, 2050014 (2020).

\bibitem{32} S. A. Franchino-Vinas , S. Mignemi, Eur. Phys. J. C \textbf{80}%
, 382 (2020).

\bibitem{33} M. Hadj Moussa, M. Merad, Few-Body Syst. \textbf{59}, 44 (2018).

\bibitem{34} M. Merad, M. Hadj Moussa, Few-Body Syst. \textbf{59}, 5 (2018).

\bibitem{35} M. Falek, M. Merad and T. Birkandan, J. Math. Phys. \textbf{58}%
, 023501 (2017).

\bibitem{36} H. Hassanabadi, E. Maghsoodi, W. S. Chung and M. de Montigny,
Eur. Phys. J. C \textbf{79}, 936 (2019).

\bibitem{37} M. M. Stetsko, J. Math. Phys. \textbf{56}, 012101 (2015).

\bibitem{g0} A. K. Geim and K. S. Novoselov, Nanoscience and Technology: A
Collection of Reviews from Nature Journals (World scientific, Singapore,
2010), pp, 11-19.

\bibitem{g1} O. L. Berman, R. Y. Kezerashvili and K. Ziegler, Phys. Rev. A
\textbf{87}, 042513 (2013).

\bibitem{g2} M. G\"unay, V. Karanikolas, R. Sahin, R. V. Ovali, A. Bek and
M. E. Tasgin, Phys. Rev. B \textbf{101}, 165412 (2020).

\bibitem{g3} M. Gullans, D. E. Chang, F. H. L. Koppens, F. J. G. de Abajo,
and M. D. Lukin, Phys. Rev. Lett. \textbf{111}, 247401 (2013).

\bibitem{g4} K. S. Novoselov, A. K. Geim, S. V. Morozov, D. Jiang, Y. Zhang,
S. V. Dubonos, I. V. Grigorieva, and A. A. Firsov, Science \textbf{306}, 666
(2004).

\bibitem{g5} Y. Zhang, J. P. Small, M. E. S. Amori, and P. Kim, Phys. Rev.
Lett. \textbf{94}, 176803 (2005).

\bibitem{g6} Z. Fang, S. Thongrattanasiri, A. Schlather, Z. Liu, L. Ma,
Y.Wang, P. M. Ajayan, P. Nordlander, N. J. Halas, and F. J. Garc\~{A}-a de
Abajo, ACS Nano \textbf{7}, 2388 (2013).

\bibitem{g7} A. H. Castro Neto, F. Guinea, N. M. R. Peres, K. S. Novoselov,
A. K. Geim, Rev. Mod. Phys. \textbf{81}, 109 (2009).

\bibitem{g8} M. A. H. Vozmediano, M. I. Katsnelson, and F. Guinea, Phys.
Rep. \textbf{496}, 109 (2010).


\bibitem{38} I. S. Gradshteyn and I. M. Ryzhik, Tables of Integrals, Series
and Products, Academic, New York (1980).

\bibitem{48} B. Hamil, M. Merad, Few-Body Syst. \textbf{60}, 36 (2019).

\bibitem{49} C. Bastos, O. Bertolami, N. Dias and J. Prata, Int. J. Mod.
Phys. A \textbf{28}, 1350064 (2013).

\bibitem{50} V. Santos, R. V. Maluf and C. A. S. Almeida, Ann. Phys. \textbf{%
349}, 402 (2014).

\bibitem{51} H. Bateman and A. Erdelyi, Higher Transcendental Functions,
Vol. 1, New York, McGraw-Hill Book Comp. (1953).

\bibitem{52} Ya.B. Zel'dovich and Yu.P. Raizer, Physics of Shock Waves and
High-Temperature Hydrodynamic Phenomena, Academic, New York (1966).

\bibitem{53} R, Hou\c{c}a and A. Jellal, Phys. Scr. \textbf{94}, 105707
(2019).

\bibitem{54} Kh. Nouicer, J. Phys. A: Math. Gen. \textbf{39}, 5125 (2006).

\bibitem{55} A. Boumali, Phys. Scr. \textbf{90}, 109501 (2015).

\bibitem{56} I. Pikovski, M. R. Vanner, M. Aspelmeyer, M. S. Kim, C.
Brukner, Nature Physics \textbf{8}, 393 (2012).
\end{thebibliography}
\end{document}